\documentclass[12pt]{article}
\def\d{\delta}

\def\l{\lambda}
\def\L{\Lambda}
\def\m{\mu}
\def\n{\nu}
\def\r{\rho}

\def\o{\omega}

\def\be{\begin{equation}}
\def\ee{\end{equation}}
\def\ba{\begin{eqnarray}}
\def\ea{\end{eqnarray}}

\newcommand{\no}{\nonumber}
\begin{document}
\renewcommand{\thefootnote}{\fnsymbol{footnote}}

\newpage
\setcounter{page}{0}
\pagestyle{empty}

\begin{center}
{\Large{\bf Gauge semi-simple extension of the Poincar\'e group\\}}
\vspace{1cm}
{\bf Dmitrij V. Soroka\footnote{E-mail: dsoroka@kipt.kharkov.ua} and 
Vyacheslav A. Soroka\footnote{E-mail: vsoroka@kipt.kharkov.ua}}
\vspace{1cm}\\
{\it Kharkov Institute of Physics and Technology,\\
1, Akademicheskaya St., 61108 Kharkov, Ukraine}\\
\vspace{1.5cm}
\end{center}
\begin{abstract}
Based on the gauge semi-simple tensor extension of the $D$-dimensional 
Poincar\'e group 
another alternative approach to the cosmological term problem is proposed.

\bigskip
\noindent
{\it PACS:} 02.20.Sv; 11.30.Cp; 11.15.-q

\medskip
\noindent
{\it Keywords:} Poincar\'e algebra, Tensor, Extension, Casimir operators,
Gauge group

\end{abstract}

\newpage
\pagestyle{plain}
\renewcommand{\thefootnote}{\arabic{footnote}}
\setcounter{footnote}0

%\sect{Introduction}

{\bf 1.\/} 
Recently the approach to the cosmological constant problem based on the tensor 
extension of the Poincar\'e algebra with the generators of the rotations 
$M_{ab}$ and translations $P_a$ \cite{bcr,sch,bh,gios1,gios2,cj,can,ss1,
dss0,dss,ss2,ss3,sorsor,bg,bg1,gkl,ggp,bgkl,l} 
\begin{eqnarray}\label{1}
[M_{ab},M_{cd}]=(g_{ad}M_{bc}+g_{bc}M_{ad})-(c\leftrightarrow d),
\end{eqnarray}
\begin{eqnarray}\label{2}
[M_{ab},P_c]=g_{bc}P_a-g_{ac}P_b,
\end{eqnarray}
\begin{eqnarray}\label{3}
[P_a,P_b]=cZ_{ab},
\end{eqnarray}
\begin{eqnarray}\label{4}
[M_{ab},Z_{cd}]=(g_{ad}Z_{bc}+g_{bc}Z_{ad})-(c\leftrightarrow d),
\end{eqnarray}
\begin{eqnarray}
[P_a,Z_{bc}]=0,\nonumber
\end{eqnarray}
\begin{eqnarray}
[Z_{ab},Z_{cd}]=0\no
\end{eqnarray}
was given by de Azcarraga, Kamimura and Lukierski \cite{akl}. Here $Z_{ab}$ 
is a tensor generator, $g_{ab}$ is a 
constant Minkovski metric and $c$ is some constant.

In this paper we present another approach to the problem based on the gauge 
semi-simple tensor extension of the $D$-dimensional Poincar\'e group which 
Lie algebra has the following form \cite{sorsor,sorsor1}:
\begin{eqnarray}\label{5}
[Z_{ab},P_c]={\L\over 3c}(g_{bc}P_a-g_{ac}P_b),
\end{eqnarray}
\begin{eqnarray}\label{6}
[Z_{ab},Z_{cd}]={\L\over 3c}[(g_{ad}Z_{bc}+g_{bc}Z_{ad})
-(c\leftrightarrow d)],
\end{eqnarray}
whereas the form of the rest permutation relation (\ref{1})-(\ref{4}) is not 
changed. $\L$ is some constant.

The Lie algebra (\ref{1})-(\ref{6}) has the following quadratic Casimir 
operator:
\begin{eqnarray}
P^aP_a+cZ^{ab}M_{ba}+{\L\over6}M^{ab}M_{ab}\mathrel{\mathop=^{\rm def}}
X_kh^{kl}X_l,\no
\end{eqnarray}
where $X_k=\{P_a, M_{ab}, Z_{ab}\}$ is a set of the generators for the Lie algebra under 
consideration (\ref{1})-(\ref{6}) and the tensor $h^{kl}$ is invariant with respect to 
the adjoint representation
\ba
h^{kl}={U^k}_m{U^l}_nh^{mn}.\no
\ea
The inverse tensor $h_{kl}$ $(h_{kl}h^{lm}={\d_k}^m)$ is invariant with respect to the 
co-adjoint representation
\ba
h_{kl}=h_{mn}{U^m}_k{U^n}_l.\no
\ea

{\bf 2.\/} 
Let us consider a gauge group corresponding to the Lie algebra (\ref{1})-(\ref{6}). To 
this end we introduce a gauge 1-form 
\ba
A=A^kX_k=dx^\m({e_\m}^aP_a+{1\over2}{\o_\m}^{ab}M_{ab}+{1\over2}{B_\m}^{ab}Z_{ab})\no
\ea
with the following gauge transformation:
\ba
A'=G^{-1}dG+G^{-1}AG,\no
\ea
where $G$ is a group element corresponding to the Lie algebra (\ref{1})-(\ref{6}). Here
$x^\m$ are space-time coordinates, ${e_\m}^a$ is a vierbein, ${\o_\m}^{ab}$ is a spin 
connection and ${B_\m}^{ab}$ is a gauge field conforming to the tensor generator $Z_{ab}$.

A contravariant vector $F^k$ of the field strength 2-form
\ba
F=F^kX_k=dA+A\wedge A={1\over2}dx^\m\wedge dx^\n F_{\m\n}\no
\ea
is transformed homogeneously under the gauge transformation
\ba
{F'}^kX_k={U^k}_lF^lX_k=G^{-1}F^kX_kG.\no
\ea
The field strength
\ba
F_{\m\n}={F_{\m\n}}^kX_k=\partial_{[\m}A_{\n]}+[A_\m,A_\n]\no
\ea
has the following decomposition:
\ba
F_{\m\n}={F_{\m\n}}^aP_a+{1\over2}{R_{\m\n}}^{ab}M_{ab}+{1\over2}{F_{\m\n}}^{ab}Z_{ab}.\no
\ea
Here
\ba
{F_{\m\n}}^a={T_{\m\n}}^a+{\L\over3c}{B_{[\m}}^{ab}e_{\n]b},\no
\ea
where
\ba
{T_{\m\n}}^a=\partial_{[\m}{e_{\n]}}^a+{\o_{[\m}}^{ab}e_{\n]b}\no 
\ea
is a torsion,
\ba
{R_{\m\n}}^{ab}=\partial_{[\m}{\o_{\n]}}^{ab}+{\o_{[\m}}^{ac}{\o_{\n]c}}^b\no 
\ea
is a curvature tensor and
\ba
{F_{\m\n}}^{ab}=\partial_{[\m}{B_{\n]}}^{ab}+{\o_{[\m}}^{c[a}{{B_{\n]}}^{b]}}_c+
{\L\over3c}{B_{[\m}}^{ca}{{B_{\n]}}^{b}}_c+c{e_{[\m}}^a{e_{\n]}}^b\no 
\ea
is a component corresponding to the tensor generator $Z_{ab}$.

An invariant Lagrangian has the following form:
\ba
L=-{e\over4}h_{kl}{F_{\m\n}}^l{F_{\r\l}}^kg^{\m\r}g^{\n\l}={e\over4}\left({1\over c}
{R_{\m\n}}^{ab}F_{\r\l;ab}+{\L\over6c^2}{F_{\m\n}}^{ab}F_{\r\l;ab}-
{F_{\m\n}}^aF_{\r\l;a}\right)g^{\m\r}g^{\n\l}.\no
\ea
Note that there exists a curious limit $c\to\infty$ 
which results in 
\ba
L\rightarrow%{c\to\infty}
{\cal L}=\left({1\over2}R+\L
-{1\over4}{T_{\m\n}}^a{T^{\m\n}}_a\right)e,\no 
\ea
where $R={R_{\m\n}}^{ab}{e_a}^\m{e_b}^\n$ is a scalar curvature,
$g^{\m\n}= g^{ab}{e_a}^\m {e_b}^\n$ is a metric tensor,
$e=\det{e_\m}^a$ is a determinant of the vierbein and $\L$ is a cosmological
constant.

{\bf 3.\/}
Thus, we have presented another alternative approach to the cosmological 
term problem within the gauge semi-simple tensor extension of the Poincar\'e 
group.

\end{document}